\pdfoutput=1
\documentclass[sigconf]{acmart}

\usepackage{graphicx}
\usepackage{natbib}
\usepackage{microtype}
\usepackage{libertine}

\copyrightyear{2024}
\acmYear{2024}
\setcopyright{rightsretained}
\acmConference[SIGIR '24]{Proceedings of the 47th International ACM SIGIR Conference on Research and Development in Information Retrieval}{July 14--18, 2024}{Washington, DC, USA}
\acmBooktitle{Proceedings of the 47th International ACM SIGIR Conference on Research and Development in Information Retrieval (SIGIR '24), July 14--18, 2024, Washington, DC, USA}
\acmDOI{10.1145/3626772.3657981}
\acmISBN{979-8-4007-0431-4/24/07}

\makeatletter
\gdef\@copyrightpermission{
    \begin{minipage}{0.3\columnwidth}
        \href{https://creativecommons.org/licenses/by/4.0/}{\includegraphics[width=0.90\textwidth]{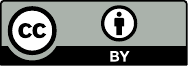}}
    \end{minipage}\hfill
    \begin{minipage}{0.7\columnwidth}
        \href{https://creativecommons.org/licenses/by/4.0/}{This work is licensed under a Creative Commons
        Attribution International 4.0 License.}
    \end{minipage}
\vspace{5pt}
}
\makeatother

\begin{document}

%\title{Towards BMI Music Information Retrieval System}

\title{Prediction of the Realisation of an Information Need: An EEG Study}


\author{Niall McGuire}
\email{niall.mcguire@strath.ac.uk}
\orcid{0009-0005-9738-047X}
\affiliation{%
   \institution{University of Strathclyde}
  %\department{Neurasearch Laboratory}
  \city{Glasgow}
  \country{United Kingdom}
}

\author{Yashar Moshfeghi}
\email{yashar.moshfeghi@strath.ac.uk}
\orcid{0000-0003-4186-1088}
\affiliation{%
  \institution{University of Strathclyde}
  %\department{Neurasearch Laboratory}
  \city{Glasgow}
  \country{United Kingdom}
}

%Change abstract
% move citations so they have a space afterwards
\begin{abstract}
%One of the foundational goals of Information Retrieval (IR) is to satisfy searchers' Information Needs (IN). Understanding how INs physically manifest has long been a complex and elusive process. However, recent studies utilising Electroencephalography (EEG) data have provided real-time insights into the neural processes associated with INs. Unfortunately, they have yet to demonstrate how this insight can practically benefit the search experience. As such, within this study, we aim to explore the efficacy of EEG data collected from 24 subjects to allow for the prediction of the realisation of an IN in real time. Additionally, we investigated the combinations of EEG features that yield optimal predictive performance and identified the specific segments within the questions where the development of subjects' INs is most pronounced. The findings from this work demonstrate that EEG data is sufficient for the real-time prediction of the realisation of an IN across all subjects with an accuracy of 73.5\% (SD 2.6\%) and on a per-subject basis with an accuracy of 90.1\% (SD 22.1\%). This work helps to close the gap by bridging theoretical neuroscientific advancements with tangible improvements in information retrieval practices, paving the way for real-time prediction of the realisation of IN. 

One of the foundational goals of Information Retrieval (IR) is to satisfy searchers' Information Needs (IN). Understanding how INs physically manifest has long been a complex and elusive process. However, recent studies utilising Electroencephalography (EEG) data have provided real-time insights into the neural processes associated with INs. Unfortunately, they have yet to demonstrate how this insight can practically benefit the search experience. As such, within this study, we explore the ability to predict the realisation of IN within EEG data across 14 subjects whilst partaking in a Question-Answering (Q/A) task. Furthermore, we investigate the combinations of EEG features that yield optimal predictive performance, as well as identify regions within the Q/A queries where a subject's realisation of IN is more pronounced. The findings from this work demonstrate that EEG data is sufficient for the real-time prediction of the realisation of an IN across all subjects with an accuracy of 73.5\% (SD 2.6\%) and on a per-subject basis with an accuracy of 90.1\% (SD 22.1\%). This work helps to close the gap by bridging theoretical neuroscientific advancements with tangible improvements in information retrieval practices, paving the way for real-time prediction of the realisation of IN. 

\end{abstract}




% %% A "teaser" image appears between the author and affiliation
% %% information and the body of the document, and typically spans the


\maketitle

\section{Introduction}
\label{introduction}
The primary objective of any Information Retrieval (IR) system is to fulfil a searcher's Information Need (IN) \cite{cooper1971definition, michalkova2022information, moshfeghi2016understanding}. In the realm of IR, numerous endeavours have been dedicated to unravelling and defining the intricate concept of IN. Pioneering works, such as Taylor's Question Negotiation Process \cite{taylor1968question}, Anomalous State of Knowledge Model \cite{belkin1982ask}, and Wilson's Information Seeking Behavior \cite{wilson1981user}, have significantly contributed to this pursuit. These works explore the essence of IN by examining user behaviour through techniques like user-system interactions \cite{bendersky2009analysis}, self-reflective notes \cite{kuhlthau1991inside}, and interviews/questionnaires \cite{markey1981levels}. While these methods offer valuable insights, reporting IN by subjects is often challenging due to its intricate and elusive nature, thereby constraining the efficacy of user-based studies \cite{moshfeghi2016understanding}.

Consequently, over the last decade, a new line of research has endeavoured to address the inherent limitations by directly examining the neurological activity in the searcher's brain through the utilisation of neuroimaging technologies\cite{moshfeghi2016understanding, moshfeghi2019neuropsychological, moshfeghi2019towards, michalkova2022information}. Research conducted at the crossroads of neuroscience and information retrieval is often referred to as \textit{NeuraSearch} \cite{moshfeghi2021neurasearch}. This interdisciplinary field has yielded numerous findings focused on the tangible representation of information needs (INs) within specific brain regions. For instance, Functional Magnetic Resonance Imaging (fMRI) was employed to observe subjects' brain activity in a Q/A task \cite{moshfeghi2016understanding, moshfeghi2019towards}. The results revealed a distributed network of brain regions commonly associated with IN, with varying activity levels in these regions based on whether the subject knew the answer to a question or needed to search for it. Further exploration \cite{moshfeghi2019towards} utilised fMRI data from a similar Q/A task to train a support vector machine (SVM) capable of distinguishing instances when a searcher possesses an IN. These investigations have presented compelling evidence regarding the existence and expression of INs within the minds of searchers. However, employing fMRI for such analyses presents several limitations. Firstly, the physical hardware of an fMRI machine is both sizable and costly, necessitating the subject to lie supine within the central bore of the apparatus while maintaining stillness, as outlined by Moshfeghi et al. \cite{moshfeghi2019towards}. Secondly, despite its fine spatial resolution, fMRI exhibits suboptimal temporal resolution, with each measurement taking a duration of 2 seconds \cite{moshfeghi2019towards}. This limitation is further exacerbated by the Blood Oxygenation Level Dependent (BOLD) signal's inherent delays \cite{moshfeghi2019towards, moshfeghi2016understanding}. Despite the valuable insights offered by fMRI, the cumbersome nature and high cost of the equipment, coupled with its temporal constraints, hinder its seamless integration into current IR systems.

Acknowledging the limitations inherent within fMRI data, researchers sought alternative neuroimaging methods to better depict the dynamics of INs with higher temporal resolution. One such approach involves the utilisation of Electroencephalography (EEG) data, a cheaper and more practical method, where electrical activity from the brain is recorded at a millisecond scale through electrodes placed on the subject's scalp \cite{teplan2002fundamentals}. In the research presented by \cite{michalkova2022information}, EEG data is employed to observe subjects' brain activity during a Q/A session. This investigation aims to understand the temporal dynamics of IN formation, detecting the presence of INs even before searchers consciously acknowledge them. This exploration opens avenues for a proactive search process, offering insights into the early stages of information needs. Although previous works \cite{michalkova2022information} provided an excellent analysis of the physical manifestations of the realisation of an IN within real-time through the use of EEG data, the question of {\it ``Can the realisation of an IN be predicted in real-time?''} is still unanswered. From this hypothesis, we formulate these four research questions: \textbf{RQ1:} "Is it possible to predict the realisation of an IN in real-time from EEG data?", \textbf{RQ2:} "Can prediction of the realisation of IN be generalised across subjects, or is it subject-specific?", \textbf{RQ3:} "During the Q/A session, where are the strongest indicators of the realisation of an IN?", \textbf{RQ4:} "What combination of features is optimal for the realisation of an IN prediction?".

In order to address our first research question, within this study we incorporated the EEG data gathered from 14 subjects whilst they took part in a Q/A task which involved the subjects observing queries word-by-word and determining if they could correctly answer the question or had a need to search (IN). This data is then provided to machine learning models to predict the subject's realisation of an IN. Additionally, we investigate the inter and intra-variability of EEG data across a variety of subjects by exploring how the prediction of the realisation of an IN is affected when the models are trained to generalise across subjects compared to when they are trained on a single subject at a time. Moreover, whilst subjects examine the queries from the Q/A tasks word-by-word, we determine which segments within the given sentences are the strongest indicator of the realisation of an IN. As well as this, we perform an ablation analysis to discern the combination of commonly extracted EEG features that enables the models to best distinguish between different search states. 

% \vspace{-15pt}

\section{Methodology}
\label{Methodology}

\noindent \textbf{subjects.} The subjects were recruited by the University of Strathclyde. They received no monetary payments but were eligible for academic credits. The subjects consisted of 13 females (93\%) and 1 male (7\%) within an age range between 18 and 39 years and a mean age of 23 years (SD 6.5).
 
% \textbf{Procedure.}
% \label{Data set}

\noindent  \textbf{Recording.} The EEG data was captured using a 40-electrode NeuroScan Ltd. system with a 10/20 cap, sampled at a frequency of 500Hz. The Q/A task was made of general knowledge questions taken from TREC-8 and TREC-2001 and B-KNorms Database\textsuperscript{2}. 
 
\noindent  \textbf{Q/A Dataset.} Two independent assessors separately evaluated the question difficulty (Cohen's Kappa: 0.61). A subset of 120 questions was then selected, and both annotators agreed upon their difficulties. The difficulty of the questions was equally distributed between easy and difficult for the overall dataset.

\begin{figure}
    \centering
    \includegraphics[width=8cm]{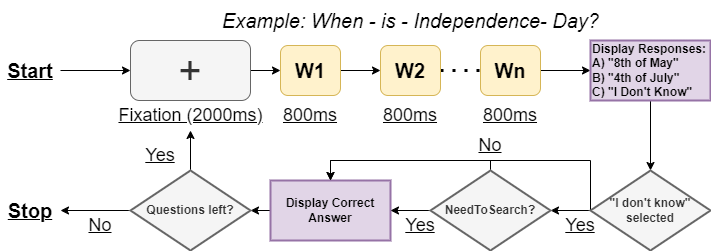}
    \caption{Task Procedure.}
    \label{fig:|Task procedure}
\end{figure}

% \subsection{Experimental Procedure}
\noindent \textbf{Experimental Procedure.}
Ethical permission to conduct the study was approved by the Universities Ethics Committee, with the tasks being conducted in a laboratory setting and all subjects meeting the inclusion criteria, i.e. healthy subjects of ages 18 - 55 years, fluent English ability, and no prior/current neurological disorders that may influence the task. Before any trials began, consent was obtained from the subjects.
To ensure the subjects had a solid grasp of the procedure, before the main trial, they were supplied with a practice example, which consisted of five questions not included in the main trial. For the practice session, there was no time limit, and subjects were allowed to repeat if required, until comfortable to proceed. The following task procedure was repeated for each trial. The trial began by viewing a fixation cross in the middle of the screen for a duration of 2000ms, indicating the location of the next stimuli on the screen, which was a way to minimise eye movements on the screen. The subjects then viewed a sequential presentation of a question randomly selected from the dataset. Each word within the question was displayed for 800ms on the screen one at a time. Within this step, the subject processed the information as it was being presented word-by-word. Following the presentation of all the words within the question, the subjects were presented with a now fully-displayed question and three on-screen answer choices associated with the question. They were requested to select the correct answer or the option "I do not know", see Figure \ref{fig:|Task procedure}. If the subjects correctly or incorrectly answered the question, the answer was displayed onscreen (NoNeedToSearch), where the trial terminated and moved on to the next question. However, if the subject selected the "I do not know", they were presented with two options: whether they wanted to search (NeedToSearch) for the correct answer or not (NoNeedToSearch). For this task, there was no search process as the overall goal was to analyse the presence of an information need based on the decision to search by the subject. After selecting one of the two options, the trial would terminate, and the next question would be presented. This was repeated for all 120 questions. Upon task completion, analysis of the 14 subjects revealed that 85\% of the responses were classed as NoNeedToSearch, and the remaining 15\% were NeedToSearch, in order to balance the dataset, the number of NoNeedToSearch classes was made equal to the number of NeedToSearch classes. The subjects completed the task (without breaks) on average in 44 min (sd=4.62, med=43.40).

% \subsection{Preprocessing} 
\noindent \textbf{Pre-processing.}
% \label{Preprocessing}
During EEG recording, the individual's actions often introduce electrical activities that can affect measurements and distort results. To address this issue, it is essential to eliminate these artefacts as effectively as possible. Initially, we utilised a bandpass filter \cite{daud2015butterworth,kingphai2021eeg} with a range of 0.5 to 50Hz. This range is commonly used because research indicates that the brain's recorded electrical activity falls within this spectrum. Additionally, we implemented average re-referencing \cite{lepage2014statistically}, a technique that establishes a reference point by aggregating the activity measured across all electrodes. The objective is to capture any noise or interference impacting all electrodes within this reference. Subsequently, we subtract this reference from each electrode's signal, effectively eliminating the noise from each electrode's signal.

% \subsection{Feature Extraction}
\noindent \textbf{Feature Extraction.}
\label{Feature Extraction}
For this study, we extract a commonly adopted core \cite{kingphai2021eeg, zhang2017classification} set of features from the EEG signals to determine which combination is optimal for IN classification, with each feature being extracted per-electrode (channel) signals across four specific frequency bands: Delta (1–4 Hz), theta (4–8 Hz), alpha (8–12 Hz), beta (12–30 Hz) and gamma (30–40 Hz). The features are extracted across every 800ms block where the question words are presented to the subject and when they respond to the question (NoNeedToSerach and NeedToSearch). The list of the features is as follows:
\textbf{Mean:} Calculates the average amplitude of the EEG signal within the specified frequency band over the 800ms block. It indicates the central tendency of the signal, helping to characterise the overall activity level. \cite{zhang2017classification}.
\textbf{Standard Deviation:} Measures the variability or spread of the EEG signal within the frequency band \cite{zhang2017classification}.
\textbf{Skewness:} Quantifies the asymmetry of the EEG signal's distribution within the frequency band \cite{kingphai2021eeg}.
\textbf{Kurtosis:} Measures the "tailedness" of the EEG signal's distribution within the frequency band \cite{kingphai2021eeg}.
\textbf{Curve Length:} Calculates the cumulative Euclidean distance between consecutive data points in the EEG signal within the frequency band \cite{zhang2017classification}.
\textbf{Number of Peaks:} Counts the number of local maxima or peaks in the EEG signal within the frequency band \cite{zhang2017classification}.
\textbf{Average Non-Linear Energy:} Quantifies the non-linear dynamics of the EEG signal within the frequency band  \cite{zhang2017classification}.

\noindent \textbf{Experiment Conditions.}
% \label{Experiment}
As the overall goal of this study is to explore the best methods for predicting IN from searchers \textbf{RQ1}, we found it key to explore several experimental parameters outlined by our research questions within Section \ref{introduction}. 

\noindent \textit{Generalised \& Personalised:} To address the research question \textbf{RQ2} during training, two methods are devised. For the first approach, the samples relating to IN and non-IN from every subject were combined into a single dataset that would then be passed onto the model, this being the generalised training strategy to asses how well our classifier can discern IN across all subjects as EEG has been noted to be heavily subject dependant. The second approach maintained the IN and non-IN EEG data at a subject level, allowing us to assess the variability of subject performance for IN prediction.

\begin{figure}
    \centering
    \includegraphics[width=7cm]{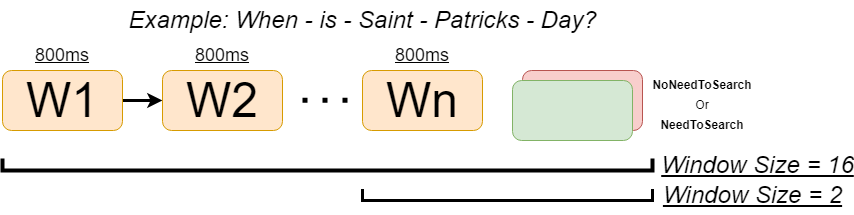}
    \caption{Expanding Window Size.}
    \label{fig:Window Size}
\end{figure}

\noindent \textit{Window Size:} To address \textbf{RQ3}, we adjusted the size of question segments (words) utilised by our classifier. This modification involved implementing an expanding window, starting from the onset of the subject's search decisions: NoNeedToSearch and NeedToSearch. On average, each question comprised seven segments, encompassing both words and responses. The minimum segment count was 4, while the maximum reached 16 segments. In this investigation, we explored the expanding window with four distinct sizes: 2, 4, 8, and 16. These sizes represented the range from the moment of question response to the full length of the question, including the response, see Figure \ref{fig:Window Size}. The objective was to ascertain the segments that the classifier favoured for distinguishing between IN and non-IN instances, potentially revealing where the realisation of IN was most pronounced during the question review process.

\noindent \textit{Feature Combination:} In accordance with \textbf{RQ4}, one of the primary aims of this study is to determine the optimal combinations of features commonly employed in EEG classification for effectively predicting the realisation of IN. As elaborated in Section \ref{Feature Extraction}, we identified and extracted seven key features for this investigation. Generating an exhaustive list of all possible combinations from these features resulted in 127 combinations. Each of these combinations was then input into the classifier, with the model's performance serving as the metric to assess the effectiveness of the various feature combinations.

% \subsection{Predictive Models}
\noindent \textbf{Predictive Models.}
For this study, we incorporated each of the aforementioned experiment conditions into a training loop, where \textit{Generalised \& Personalised} were separated where each would iterate through every possible \textit{Feature Combination} and \textit{Window Size}. The classifiers selected for this task were the Support Vector Machine (SVM) \cite{708428}, Random Forest Classifier \cite{breiman2001random}, and AdaBoost \cite{freund1996experiments} models, as they have seen substantial success within the realm of EEG classification \cite{hu2017automated, li2013feature} and are well suited to the limited quantity of data available for this task. Prior to this investigate several Deep and Recurrent Neural Networks were trained on the collected EEG data, however, their performance was sub-optimal as they were limited by the number of samples within the dataset. Each dataset provided to the model through each of the possible combinations of experiment conditions was cross-validated with a k-fold size = 5. Each fold returned the following metrics: \textit{Accuracy}, \textit{Precision}, and \textit{Recall}, where their average across each fold was calculated along with their standard deviation. \textbf{Baseline.} Since there are no prior works to compare to, we introduce a baseline that represents an untrained model where all its predictions are based on a random choice, i.e. where its accuracy is set to 50\%.

\section{Results \& Conclusion}

The results produced for the \textit{Generalised} and \textit{Personalised} conditions are detailed in Table \ref{tab:generalised} and \ref{tab:personalised} respectively. Each of these tables denotes the Model, the selected window size (W-Size), and the best-performing feature combination at the given window size with its subsequent Accuracy, Precision, and Recall scores. We also performed a paired Wilcoxon test between the predictions obtained for each model to check the significance of the difference with the baseline. All of the results obtained from our models trained on a set of features were different from that of the baseline with a confidence level of (p < 0.01).

\begin{table}[h]
\caption{\scriptsize This table shows the prediction accuracy of our Generalised approach. The standard deviation is presented in parentheses. The best-performing model is highlighted in bold.}
\centering
\resizebox{\columnwidth}{!}{%
\begin{tabular}{l|l|l|l|l|l}
Model & W-Size & Features & Accuracy (SD) & Precision (SD) & Recall (SD) \\ \hline
Baseline (Random)& - & - & 50\% & 50\% & 50\% \\ \hline

\textbf{RandomForest} & \textbf{2} & \textbf{Mean-SD-Curve} & \textbf{73.5\% (2.6\%)} & \textbf{73.6\% (2.5\%)} & \textbf{73.5\% (2.6\%)} \\
 & 4 & Mean-SD-Curve & 71.9\% (2.3\%) & 72.0\% (2.2\%) & 71.9\% (2.3\%) \\
 & 8 & Mean-SD-Curve-AvEn & 71.8\% (0.9\%) & 71.8\% (0.9\%) & 71.8\% (0.9\%) \\
 & 16 & Mean-Curve-AvEn & 71.5\% (1.1\%) & 71.6\% (1.1\%) & 71.5\% (1.1\%) \\ \hline
SVM & 2 & Mean-AvEn & 69.7\% (2.1\%) & 69.8\% (2.2\%) & 69.7\% (2.1\%) \\
 & 4 & Mean-AvEn & 69.1\% (2.4\%) & 69.2\% (2.4\%) & 69.1\% (2.4\%) \\
 & 8 & Mean-Skew-AvEn & 69.0\% (1.0\%) & 69.0\% (1.0\%) & 69.0\% (1.0\%) \\
 & 16 & Mean-Peaks & 69.0\% (1.1\%) & 69.1\% (1.1\%) & 69.0\% (1.1\%) \\ \hline
AdaBoost & 2 & Mean-SD & 70.6\% (0.2\%) & 70.7\% (0.2\%) & 70.6\% (0.2\%) \\
 & 4 & Mean-Curve & 70.4\% (0.2\%) & 70.0\% (0.1\%) & 69.9\% (0.2\%) \\
 & 8 & Mean-Curve & 70.3\% (0.1\%) & 70.5\% (0.1\%) & 70.3\% (0.1\%) \\
 & 16 & Mean-Curve-Peaks & 70.0\% (0.1\%) & 70.2\% (0.1\%) & 70.0\% (0.1\%)
\end{tabular}%
}
\label{tab:generalised}
\end{table}

% \usepackage{graphicx}
% \usepackage{graphicx}
% \usepackage{graphicx}
\begin{table}[h]
\caption{\scriptsize This table shows the prediction accuracy of our Personalised approach. The standard deviation is presented in parentheses. The best-performing model is highlighted in bold.}
\centering
\resizebox{\columnwidth}{!}{%
\begin{tabular}{l|l|l|l|l|l}
Model & W-Size & Features & Accuracy (SD) & Precision (SD) & Recall (SD) \\ \hline
Baseline (Random)& - & - & 50\% & 50\% & 50\% \\ \hline

RandomForest & 2 & Mean-SD-AvEn & 68.9\% (19.7\%) & 71.4\% (20.3\%) & 68.9\% (19.7\%) \\
 & 4 & Mean & 73.0\% (21.5\%) & 75.7\% (21.9\%) & 73.0\% (21.5\%) \\
 & 8 & Mean & 75.9\% (21.5\%) & 77.9\% (21.5\%) & 75.9\% (21.5\%) \\
 & 16 & Mean & 76.6\% (21.3\%) & 77.5\% (21.4\%) & 76.6\% (21.3\%) \\ \hline
SVM & 2 & Mean-Curver-AvEn & 74.2\% (21.5\%) & 75.4\% (21.3\%) & 74.2\% (21.5\%) \\
 & 4 & Mean-Kur-Curve & 74.3\% (21.9\%) & 76.5\% (22.1\%) & 74.3 (21.9\%) \\
 & 8 & Mean-AvEn & 73.5\% (21.8\%) & 75.7\% (22.3\%) & 73.5\% (21.8\%) \\
 & 16 & Mean-Kur-Curve-AvEn & 71.7\% (22.1\%) & 74.6\% (22.1\%) & 71.7\% (22.1\%) \\ \hline
\textbf{AdaBoost} & 2 & Mean-Peaks & 73.9\% (23.4\%) & 74.5\% (23.9\%) & 73.9\% (23.4\%) \\
 & 4 & Mean & 80.5\% (22.3\%) & 81.3\% (22.4\%) & 80.5\% (22.3\%) \\
 & 8 & Mean-Peaks & 88.9\% (22.5\%) & 89.0\% (22.2\%) & 88.9\% (22.5\%) \\
 & \textbf{16} & \textbf{Mean} & \textbf{90.1\% (22.1\%)} & \textbf{90.3\% (22.1\%)} & \textbf{90.1\% (22.1\%)}
\end{tabular}%
}
\label{tab:personalised}
\end{table}

We first address \textbf{RQ1} by reviewing the results produced in both the \textit{Generalised} and \textit{Personalised} conditions. As we can see the prediction of the realisation of IN is possible, as every model in Table \ref{tab:generalised} and Table \ref{tab:personalised} was able to achieve an accuracy score above that of random classification (50\%), with the lowest reported accuracy score being the RandomForest classifier with an accuracy of 68.9\% (SD 19.7\%) and the highest being the AdaBoost model with a score of 90.1\% (22.1\%) as seen in Table \ref{tab:personalised}. These results demonstrate that EEG data is capable of achieving greater \textit{Generalised} and \textit{Perosnalised} accuracy performance for the prediction of the realisation of IN than that of alternative neuroimaging techniques such as fMRI \cite{moshfeghi2019towards}.

By comparing the performance of the \textit{Genralised} and \textit{Personalised} models, it can be observed that the \textit{Personalised} approach achieves the highest overall prediction accuracy, evidenced by the AdaBoost model that obtained 90.1\% (SD 22.1\%) in Table \ref{tab:personalised}. When trained using the \textit{Personalised} method, the RandomForest, SVM, and AdaBoost model's accuracy on average across window sizes increases over its \textit{Generalised} counterparts by 1.4\%, 4.2\%, and 13\% respectively. However, this increased accuracy also comes with an increased Standard Deviation, with the \textit{Personalsied} RandomForest, SVM, and AdaBoost models on average across window sizes having a higher Standard Deviation of 19.3\%, 21.8\%, and 22.4\% respectively than the \textit{Generalised} models. These findings help to address \textbf{RQ2} as they suggest, on average, creating a model tailored to each subject is the best approach for predicting the realisation of IN as evidenced by the performance of the AdaBoost model in Table \ref{tab:personalised}. However, the variation in Standard deviation indicates that the \textit{Generalised} models offer a more robust and reliable prediction accuracy. This difference follows the trend observed in prior works \cite{moshfeghi2019towards} and is likely due to the natural variability in EEG data collected across subjects. As such, for future systems aiming to predict the realisation of an IN from EEG, the best approach may be to assess the model performance on individual subjects and determine if the trade-off between accuracy and standard deviation is acceptable or if a generalised model with a lower accuracy but more reliable standard deviation is more suitable for their specific research purposes.

Regarding \textbf{RQ3}, the results presented in Table \ref{tab:generalised} and Table \ref{tab:personalised} are in contrast to each other. In the \textit{Generalised} condition Table \ref{tab:generalised}, we observe that all models achieve their peak performance when the window size is set to 2, with the RandomForest, SVM and AdaBoost models achieving an accuracy of 73.5\%, 69.7\%, and 70.6\% respectively. As the window size is increased from 2 up to 16, the performance of the RandomForest, SVM, and AdaBoost models decreases by 2\%, 0.7\%, and 0.6\% respectively. Conversely, in the \textit{Personalised} results, Table \ref{tab:personalised} we observe that at the window size of 16, the RandomForest and AdaBoost models achieve their highest performance of 76.6\% and 90.1\% respectively. As the window size is increased from 2 up to 16 the RandomForest and AdaBoost models accuracy increases by 7.7\% and 16.2\% respectively, except the SVM model, which follows the same trend as the \textit{Generalised} Models.

The results produced by the \textit{Generalised} approach suggest that the distinctive EEG patterns associated with the realisation of an IN may be more strongly concentrated immediately after the subject concludes their review of the question. This might be indicative of a universal or commonly shared cognitive response that occurs promptly after the comprehension of a question, highlighting a quick and standardised recognition process for INs across subjects. In Contrast to this, the performance of the \textit{Personalsied} models indicates that for individual subjects, the discernible EEG patterns unfold over a more extended period. This could be influenced by varying cognitive styles, attention spans, or information processing speeds unique to each subject. subjects might take more time to process and formulate their information needs, leading to a prolonged period of activity associated with information-seeking. Lastly, addressing \textbf{RQ4}, we can observe that the best-performing feature is the Mean value taken from the EEG segments, as it appeared in every single best-performing combination at each window size for both \textit{generalised} and \textit{personalised} training Table \ref{tab:generalised} and \ref{tab:personalised} respectively. However, a large subset of key features see substantial use across both training conditions, for \textit{Generalised} condition the following features are listed in order of occurrence: with Curve Length appearing in 7 optimal combinations, Average Energy in 5, Standard Deviation in 4, Number of Peaks in 2, and Skewness in 1. Similarly for \textit{Personalised}: Average Energy appears in 4, Curve Length in 3, Kurtosis in 2, and Standard Deviation appears in one. Our results indicate that these features are strong performers in predicting the realisation of IN. 

% \vspace{-10pt}

% \section{Conclusion}
In conclusion, the findings of this study demonstrate that through the use of Electroencephalography (EEG) data, we were able to predict the realisation of IN substantially above the random baseline classification accuracy of 50\%, with models achieving up to 90.1\% accuracy. This work is the first to ever demonstrate the prediction of the realisation of IN through the use of EEG data, and at an accuracy higher than any other previously utilised neuroimaging techniques, paving the way to real-time realisation of IN prediction. Furthermore, the encouraging results obtained from the \textit{Generalised} and \textit{Personalised} conditions will help to inform future research and Information Retrieval (IR) systems that seek to incorporate the realisation of IN prediction, by taking into consideration the inter and intra-variability of EEG data cross subjects and examine the trade-off between a potentially more accurate subject-specific models and a more reliable generalised model. Moreover, we also highlighted optimal ranges within queries that should be examined to provide the strongest indicators of the realisation of IN, as well as the optimal combinations of features that should be considered for the prediction of the realisation of IN within subjects.

\begin{acks}
This work was supported by the Engineering and Physical Sciences Research Council [grant number EP/W522260/1].
\end{acks}

\bibliographystyle{ACM-Reference-Format}
\bibliography{ref}


\begin{thebibliography}{22}

%%% the Web address.

\ifx \showCODEN    \undefined \def \showCODEN     #1{\unskip}     \fi
\ifx \showDOI      \undefined \def \showDOI       #1{#1}\fi
\ifx \showISBNx    \undefined \def \showISBNx     #1{\unskip}     \fi
\ifx \showISBNxiii \undefined \def \showISBNxiii  #1{\unskip}     \fi
\ifx \showISSN     \undefined \def \showISSN      #1{\unskip}     \fi
\ifx \showLCCN     \undefined \def \showLCCN      #1{\unskip}     \fi
\ifx \shownote     \undefined \def \shownote      #1{#1}          \fi
\ifx \showarticletitle \undefined \def \showarticletitle #1{#1}   \fi
\ifx \showURL      \undefined \def \showURL       {\relax}        \fi
% invisible to TeX
\providecommand\bibfield[2]{#2}
\providecommand\bibinfo[2]{#2}
\providecommand\natexlab[1]{#1}
\providecommand\showeprint[2][]{arXiv:#2}

\bibitem[Belkin et~al\mbox{.}(1982)]%
        {belkin1982ask}
\bibfield{author}{\bibinfo{person}{Nicholas~J Belkin}, \bibinfo{person}{Robert~N Oddy}, {and} \bibinfo{person}{Helen~M Brooks}.} \bibinfo{year}{1982}\natexlab{}.
\newblock \showarticletitle{ASK for information retrieval: Part I. Background and theory}.
\newblock \bibinfo{journal}{\emph{Journal of documentation}} \bibinfo{volume}{38}, \bibinfo{number}{2} (\bibinfo{year}{1982}), \bibinfo{pages}{61--71}.
\newblock


\bibitem[Bendersky and Croft(2009)]%
        {bendersky2009analysis}
\bibfield{author}{\bibinfo{person}{Michael Bendersky} {and} \bibinfo{person}{W~Bruce Croft}.} \bibinfo{year}{2009}\natexlab{}.
\newblock \showarticletitle{Analysis of long queries in a large scale search log}. In \bibinfo{booktitle}{\emph{Proceedings of the 2009 workshop on Web Search Click Data}}. \bibinfo{pages}{8--14}.
\newblock


\bibitem[Breiman(2001)]%
        {breiman2001random}
\bibfield{author}{\bibinfo{person}{Leo Breiman}.} \bibinfo{year}{2001}\natexlab{}.
\newblock \showarticletitle{Random forests}.
\newblock \bibinfo{journal}{\emph{Machine learning}}  \bibinfo{volume}{45} (\bibinfo{year}{2001}), \bibinfo{pages}{5--32}.
\newblock


\bibitem[Cooper(1971)]%
        {cooper1971definition}
\bibfield{author}{\bibinfo{person}{William~S Cooper}.} \bibinfo{year}{1971}\natexlab{}.
\newblock \showarticletitle{A definition of relevance for information retrieval}.
\newblock \bibinfo{journal}{\emph{Information storage and retrieval}} \bibinfo{volume}{7}, \bibinfo{number}{1} (\bibinfo{year}{1971}), \bibinfo{pages}{19--37}.
\newblock


\bibitem[Daud and Sudirman(2015)]%
        {daud2015butterworth}
\bibfield{author}{\bibinfo{person}{SS Daud} {and} \bibinfo{person}{R Sudirman}.} \bibinfo{year}{2015}\natexlab{}.
\newblock \showarticletitle{Butterworth bandpass and stationary wavelet transform filter comparison for electroencephalography signal}. In \bibinfo{booktitle}{\emph{2015 6th international conference on intelligent systems, modelling and simulation}}. IEEE, \bibinfo{pages}{123--126}.
\newblock


\bibitem[Freund et~al\mbox{.}(1996)]%
        {freund1996experiments}
\bibfield{author}{\bibinfo{person}{Yoav Freund}, \bibinfo{person}{Robert~E Schapire}, {et~al\mbox{.}}} \bibinfo{year}{1996}\natexlab{}.
\newblock \showarticletitle{Experiments with a new boosting algorithm}. In \bibinfo{booktitle}{\emph{icml}}, Vol.~\bibinfo{volume}{96}. Citeseer, \bibinfo{pages}{148--156}.
\newblock


\bibitem[Hearst et~al\mbox{.}(1998)]%
        {708428}
\bibfield{author}{\bibinfo{person}{M.A. Hearst}, \bibinfo{person}{S.T. Dumais}, \bibinfo{person}{E. Osuna}, \bibinfo{person}{J. Platt}, {and} \bibinfo{person}{B. Scholkopf}.} \bibinfo{year}{1998}\natexlab{}.
\newblock \showarticletitle{Support vector machines}.
\newblock \bibinfo{journal}{\emph{IEEE Intelligent Systems and their Applications}} \bibinfo{volume}{13}, \bibinfo{number}{4} (\bibinfo{year}{1998}), \bibinfo{pages}{18--28}.
\newblock
\urldef\tempurl%
\url{https://doi.org/10.1109/5254.708428}
\showDOI{\tempurl}


\bibitem[Hu(2017)]%
        {hu2017automated}
\bibfield{author}{\bibinfo{person}{Jianfeng Hu}.} \bibinfo{year}{2017}\natexlab{}.
\newblock \showarticletitle{Automated detection of driver fatigue based on AdaBoost classifier with EEG signals}.
\newblock \bibinfo{journal}{\emph{Frontiers in computational neuroscience}}  \bibinfo{volume}{11} (\bibinfo{year}{2017}), \bibinfo{pages}{72}.
\newblock


\bibitem[Kingphai and Moshfeghi(2021)]%
        {kingphai2021eeg}
\bibfield{author}{\bibinfo{person}{Kunjira Kingphai} {and} \bibinfo{person}{Yashar Moshfeghi}.} \bibinfo{year}{2021}\natexlab{}.
\newblock \showarticletitle{On EEG preprocessing role in deep learning effectiveness for mental workload classification}. In \bibinfo{booktitle}{\emph{Human Mental Workload: Models and Applications: 5th International Symposium, H-WORKLOAD 2021, Virtual Event, November 24--26, 2021, Proceedings 5}}. Springer, \bibinfo{pages}{81--98}.
\newblock


\bibitem[Kuhlthau(1991)]%
        {kuhlthau1991inside}
\bibfield{author}{\bibinfo{person}{Carol~C Kuhlthau}.} \bibinfo{year}{1991}\natexlab{}.
\newblock \showarticletitle{Inside the search process: Information seeking from the user's perspective}.
\newblock \bibinfo{journal}{\emph{Journal of the American society for information science}} \bibinfo{volume}{42}, \bibinfo{number}{5} (\bibinfo{year}{1991}), \bibinfo{pages}{361--371}.
\newblock


\bibitem[Lepage et~al\mbox{.}(2014)]%
        {lepage2014statistically}
\bibfield{author}{\bibinfo{person}{Kyle~Q Lepage}, \bibinfo{person}{Mark~A Kramer}, {and} \bibinfo{person}{Catherine~J Chu}.} \bibinfo{year}{2014}\natexlab{}.
\newblock \showarticletitle{A statistically robust EEG re-referencing procedure to mitigate reference effect}.
\newblock \bibinfo{journal}{\emph{Journal of neuroscience methods}}  \bibinfo{volume}{235} (\bibinfo{year}{2014}), \bibinfo{pages}{101--116}.
\newblock


\bibitem[Li et~al\mbox{.}(2013)]%
        {li2013feature}
\bibfield{author}{\bibinfo{person}{Shufang Li}, \bibinfo{person}{Weidong Zhou}, \bibinfo{person}{Qi Yuan}, \bibinfo{person}{Shujuan Geng}, {and} \bibinfo{person}{Dongmei Cai}.} \bibinfo{year}{2013}\natexlab{}.
\newblock \showarticletitle{Feature extraction and recognition of ictal EEG using EMD and SVM}.
\newblock \bibinfo{journal}{\emph{Computers in biology and medicine}} \bibinfo{volume}{43}, \bibinfo{number}{7} (\bibinfo{year}{2013}), \bibinfo{pages}{807--816}.
\newblock


\bibitem[Markey(1981)]%
        {markey1981levels}
\bibfield{author}{\bibinfo{person}{Karen Markey}.} \bibinfo{year}{1981}\natexlab{}.
\newblock \showarticletitle{Levels of question formulation in negotiation of information need during the online presearch interview: A proposed model}.
\newblock \bibinfo{journal}{\emph{Information Processing \& Management}} \bibinfo{volume}{17}, \bibinfo{number}{5} (\bibinfo{year}{1981}), \bibinfo{pages}{215--225}.
\newblock


\bibitem[Michalkova et~al\mbox{.}(2022)]%
        {michalkova2022information}
\bibfield{author}{\bibinfo{person}{Dominika Michalkova}, \bibinfo{person}{Mario Parra-Rodriguez}, {and} \bibinfo{person}{Yashar Moshfeghi}.} \bibinfo{year}{2022}\natexlab{}.
\newblock \showarticletitle{Information Need Awareness: an EEG study}. In \bibinfo{booktitle}{\emph{Proceedings of the 45th International ACM SIGIR Conference on Research and Development in Information Retrieval}}. \bibinfo{pages}{610--621}.
\newblock


\bibitem[Moshfeghi(2021)]%
        {moshfeghi2021neurasearch}
\bibfield{author}{\bibinfo{person}{Yashar Moshfeghi}.} \bibinfo{year}{2021}\natexlab{}.
\newblock \showarticletitle{Neurasearch: Neuroscience and information retrieval}. In \bibinfo{booktitle}{\emph{CEUR Workshop Proceedings}}, Vol.~\bibinfo{volume}{2950}. \bibinfo{pages}{193--194}.
\newblock


\bibitem[Moshfeghi and Pollick(2019)]%
        {moshfeghi2019neuropsychological}
\bibfield{author}{\bibinfo{person}{Yashar Moshfeghi} {and} \bibinfo{person}{Frank~E Pollick}.} \bibinfo{year}{2019}\natexlab{}.
\newblock \showarticletitle{Neuropsychological model of the realization of information need}.
\newblock \bibinfo{journal}{\emph{Journal of the Association for Information Science and Technology}} \bibinfo{volume}{70}, \bibinfo{number}{9} (\bibinfo{year}{2019}), \bibinfo{pages}{954--967}.
\newblock


\bibitem[Moshfeghi et~al\mbox{.}(2019)]%
        {moshfeghi2019towards}
\bibfield{author}{\bibinfo{person}{Yashar Moshfeghi}, \bibinfo{person}{Peter Triantafillou}, {and} \bibinfo{person}{Frank Pollick}.} \bibinfo{year}{2019}\natexlab{}.
\newblock \showarticletitle{Towards predicting a realisation of an information need based on brain signals}. In \bibinfo{booktitle}{\emph{The world wide web conference}}. \bibinfo{pages}{1300--1309}.
\newblock


\bibitem[Moshfeghi et~al\mbox{.}(2016)]%
        {moshfeghi2016understanding}
\bibfield{author}{\bibinfo{person}{Yashar Moshfeghi}, \bibinfo{person}{Peter Triantafillou}, {and} \bibinfo{person}{Frank~E Pollick}.} \bibinfo{year}{2016}\natexlab{}.
\newblock \showarticletitle{Understanding information need: An fMRI study}. In \bibinfo{booktitle}{\emph{Proceedings of the 39th International ACM SIGIR conference on Research and Development in Information Retrieval}}. \bibinfo{pages}{335--344}.
\newblock


\bibitem[Taylor(1968)]%
        {taylor1968question}
\bibfield{author}{\bibinfo{person}{Robert~S Taylor}.} \bibinfo{year}{1968}\natexlab{}.
\newblock \showarticletitle{Question-negotiation and information seeking in libraries}.
\newblock \bibinfo{journal}{\emph{College \& research libraries}} \bibinfo{volume}{29}, \bibinfo{number}{3} (\bibinfo{year}{1968}), \bibinfo{pages}{178--194}.
\newblock


\bibitem[Teplan et~al\mbox{.}(2002)]%
        {teplan2002fundamentals}
\bibfield{author}{\bibinfo{person}{Michal Teplan} {et~al\mbox{.}}} \bibinfo{year}{2002}\natexlab{}.
\newblock \showarticletitle{Fundamentals of EEG measurement}.
\newblock \bibinfo{journal}{\emph{Measurement science review}} \bibinfo{volume}{2}, \bibinfo{number}{2} (\bibinfo{year}{2002}), \bibinfo{pages}{1--11}.
\newblock


\bibitem[Wilson(1981)]%
        {wilson1981user}
\bibfield{author}{\bibinfo{person}{Tom~D Wilson}.} \bibinfo{year}{1981}\natexlab{}.
\newblock \showarticletitle{On user studies and information needs}.
\newblock \bibinfo{journal}{\emph{Journal of documentation}} \bibinfo{volume}{37}, \bibinfo{number}{1} (\bibinfo{year}{1981}), \bibinfo{pages}{3--15}.
\newblock


\bibitem[Zhang et~al\mbox{.}(2017)]%
        {zhang2017classification}
\bibfield{author}{\bibinfo{person}{Yong Zhang}, \bibinfo{person}{Bo Liu}, \bibinfo{person}{Xiaomin Ji}, {and} \bibinfo{person}{Dan Huang}.} \bibinfo{year}{2017}\natexlab{}.
\newblock \showarticletitle{Classification of EEG signals based on autoregressive model and wavelet packet decomposition}.
\newblock \bibinfo{journal}{\emph{Neural Processing Letters}}  \bibinfo{volume}{45} (\bibinfo{year}{2017}), \bibinfo{pages}{365--378}.
\newblock


\end{thebibliography}
\end{document}